%% file: isotopologues.tex
\pdfoutput=1
\documentclass[twocolumn,times]{aastex62} 
\usepackage{graphicx}
\usepackage{mathtools,upgreek} 
\usepackage{natbib}
\usepackage{outlines}
\usepackage{amsmath}
\usepackage{verbatim}
\usepackage{appendix}
\usepackage{multirow,tabularx}
\usepackage[version=4]{mhchem} 

\newcommand{\um}{$\upmu$m}
\newcommand{\OCO}{\ce{^18O^12C^16O}}
\newcommand{\SN}{$\langle$S/N$\rangle$}

\received{November 20, 2018}
\revised{May 7, 2019} 
\accepted{May 22, 2019}
\submitjournal{The Astronomical Journal}

\newif\iftwocol
\twocoltrue

\begin{document}

\title{{Observing Isotopologue Bands in Terrestrial Exoplanet Atmospheres with the \textit{James Webb Space Telescope}---\\
Implications for Identifying Past Atmospheric and Ocean Loss}}

\correspondingauthor{Andrew P. Lincowski}
\email{alinc@uw.edu}

\author[0000-0003-0429-9487]{Andrew P. Lincowski}
\affiliation{Department of Astronomy and Astrobiology Program, University of Washington, Box 351580, Seattle, Washington 98195, USA}
\affiliation{NASA NExSS Virtual Planetary Laboratory, Box 351580, University of Washington, Seattle, Washington 98195, USA}

\author[0000-0002-0746-1980]{Jacob Lustig-Yaeger}
\affiliation{Department of Astronomy and Astrobiology Program, University of Washington, Box 351580, Seattle, Washington 98195, USA}
\affiliation{NASA NExSS Virtual Planetary Laboratory, Box 351580, University of Washington, Seattle, Washington 98195, USA}

\author[0000-0002-1386-1710]{Victoria S. Meadows}
\affiliation{Department of Astronomy and Astrobiology Program, University of Washington, Box 351580, Seattle, Washington 98195, USA}
\affiliation{NASA NExSS Virtual Planetary Laboratory, Box 351580, University of Washington, Seattle, Washington 98195, USA}

\shorttitle{Observing Isotopologue Bands with JWST} 

\shortauthors{Lincowski, Lustig-Yaeger, \& Meadows}

\begin{abstract}
Terrestrial planets orbiting M~dwarfs may soon be observed with the \textit{James Webb Space Telescope} (\textit{JWST}) to characterize their atmospheric composition and search for signs of habitability or life.  These planets may undergo significant atmospheric and ocean loss due to the superluminous pre-main-sequence phase of their host stars, which may leave behind abiotically-generated oxygen, a false positive for the detection of life. Determining if ocean loss has occurred will help assess potential habitability and whether or not any \ce{O2} detected is biogenic. In the solar system, differences in isotopic abundances have been used to infer the history of ocean loss and atmospheric escape (e.g. Venus, Mars). We find that isotopologue measurements using transit transmission spectra of terrestrial planets around late-type M~dwarfs like TRAPPIST-1 may be possible with \textit{JWST}, if the escape mechanisms and resulting isotopic fractionation were {similar to} Venus. We present analyses of post-ocean-loss \ce{O2}- and \ce{CO2}-dominated atmospheres, containing a range of trace gas abundances. Isotopologue bands are likely detectable throughout the near-infrared (1--8~\um{}), especially 3--4~\um{}, although not in \ce{CO2}-dominated atmospheres.  For Venus-like D/H {ratios }100 times that of Earth, TRAPPIST-1~b transit signals of up to {79}~ppm are possible by observing HDO. Similarly, \ce{^18O/^16O} {ratios} 100 times that of Earth {produce signals} at up to 94~ppm. Detection at S/N=5 may be attained on these bands with as few as four to {eleven} transits, with optimal use of \textit{JWST} NIRSpec Prism. Consequently, \ce{H2O} and \ce{CO2} isotopologues could be considered as indicators of {past} ocean loss and atmospheric escape for \textit{JWST} observations of terrestrial planets around M~dwarfs.
\end{abstract}


\section{Introduction}

In the near future, terrestrial exoplanets around small M~dwarf stars will be observed by the \textit{James Webb Space Telescope} (\textit{JWST}) and extremely large ground-based telescopes \citep{Cowan:2015,Quanz:2015,Snellen:2015,Greene:2016,Lovis:2017,Morley:2017,Lincowski:2018}. A number of nearby transiting targets have been discovered, including the seven-planet TRAPPIST-1 system \citep{Gillon:2016,Gillon:2017,Luger:2017b}, which provides a plausible opportunity for studying the liquid water habitable zone {(HZ)} and planetary evolution in a single system. However, the atmospheres of these planets will likely be heavily evolved from their primordial composition due to the long, superluminous pre-main-sequence evolution \citep{Baraffe:2015} and life-long stellar activity \citep{Tarter:2007} of their M~dwarf host stars.

The superluminous pre-main-sequence phase of M~dwarf stars may drive significant loss of a planet's surface water and atmosphere, and potentially produce large quantities of atmospheric oxygen. This superluminous phase could last for up to one billion years for the smallest stars  \citep{Baraffe:2015}. During this time, ocean-bearing planets that formed in what is presently the habitable zone would have been subjected to fluxes up to 100 times the {stellar irradiation of the main-sequence. This} would cause a runaway greenhouse environment and severe hydrodynamic escape \citep{Luger:2015b}{, which occurs when heating by extreme UV absorption induces an escape flow \citep[e.g.][]{Hunten:1987}}. During this period, the TRAPPIST-1 planets could have lost up to twenty Earth oceans of water and generated thousands of bars of oxygen \citep[c.f.][]{Bolmont:2017,Wordsworth:2018,Lincowski:2018}. This oxygen could remain in the atmosphere, but is more likely to be severely reduced by a number of planetary processes.  These processes include oxidation of the surface, interaction with a magma ocean that reincorporates the \ce{O2} into the mantle \citep{Schaefer:2016,Wordsworth:2018}, or loss of \ce{O2} to space either via hydrodynamic escape early on or by a number of ongoing escape mechanisms \citep[e.g.][]{Hunten:1982,Lammer:2007,Ribas:2016,Airapetian:2017,Dong:2017,Garcia:2017,Egan:2019}.

Any atmospheric oxygen left by these {sequestration/loss} processes could constitute a false positive biosignature if the planet is being assessed for signs of life \citep{Meadows:2017,Meadows:2018c}. {Even with sequestration and loss processes, ocean loss may leave behind several bars of \ce{O2} \citep{Schaefer:2016, Wordsworth:2018}, which could be potentially discriminated from the modest amounts generated by a photosynthetic biosphere via detection of \ce{O2}-\ce{O2} collision-induced absorption bands. The \ce{O2}-\ce{O2} bands, particularly at 1.06 and 1.27~\um{}, are more prominent in massive \ce{O2} atmospheres because the absorption cross-section is proportional to the density of gas squared \citep{Schwieterman:2016,Meadows:2018a,Lincowski:2018}. Detection of these bands alone does not prove that the planet lacks a surface ocean, only that large amounts of oxygen have been liberated, and this is likely from water vapor photolysis and subsequent hydrogen escape.}

Evidence of {past} atmospheric escape and ocean loss may help test the concept of the habitable zone, that region around a star where liquid water could exist on the surface of an Earth-like planet \citep{Kasting:1993,Kopparapu:2013}. The inner edge of the habitable zone (IHZ) is conservatively defined by the moist greenhouse greenhouse limit (where stratospheric \ce{H2O} exceeds 1000~ppm and so water {vapor} is easily lost to space), which lies close to Earth (0.99~au around a Sun-like star). However, an optimistic ``recent Venus'' limit can be defined under the assumption that Venus may have had surface water{ prior to approximately one billion years ago, before the last global resurfacing event, when the Sun was fainter \citep{Solomon:1991,Kasting:1993}}. Additionally, there have been a number of modeling studies positing revised limits for the IHZ that depend on perturbations to some of the parameter assumptions, including planetary mass \citep{Kopparapu:2014} and rotation rates \citep[e.g.][]{Kopparapu:2016}. Since TRAPPIST-1~d is between the {recent} Venus limit and the conservative inner edge as defined by \citet{Kopparapu:2013}, it is a valuable target for probing the {position of the }inner edge of the habitable zone around M~dwarf stars.

In a multi-planet system like TRAPPIST-1, evidence of {past atmospheric escape from} the inner planet(s) could inform the suitability for more difficult follow-up observations of a habitable zone target (e.g. TRAPPIST-1~e). {It is much easier to characterize the inner planets due to the possibility of obtaining more transit observations and the larger scale heights afforded by the hotter atmospheres \citep{Morley:2017,Lincowski:2018}. The survival of an atmosphere inward of the IHZ could be an indicator for atmospheric survival of the other planets. For example, if TRAPPIST-1~b still has an atmosphere, then planets farther away---including those in the habitable zone---have a higher likelihood of also hosting an atmosphere. However, the presence of an atmosphere on a habitable zone planet does not guarantee the planet is habitable, as even habitable zone planets may have undergone complete ocean loss \citep{Lincowski:2018}.}

{Another piece of evidence that }can indicate a planetary environment lost its surface water, so is not likely to be habitable{, is }severe isotopic fractionation. Both Venus and Mars once likely had surface oceans \citep[e.g.][]{deBergh:1991,Wordsworth:2016}, {evidence for which includes} isotopic fractionation in the observed {atmospheric }deuterium to hydrogen ratios (D/H) compared to Vienna Standard Mean Ocean Water (VSMOW) for Earth. {Compared to VSMOW}, the atmosphere of Venus is enhanced by a factor of 120--140 \citep{deBergh:1991,Matsui:2012} and the atmosphere of Mars is enhanced by a factor of $\sim$4 \citep{Owen:1988,Villanueva:2015,Encrenaz:2018}. These enhancements likely occurred from near-complete loss of their available water reservoirs \citep{Hunten:1982,Owen:1988}, because any primordial reservoir would dilute fractionated gas and reduce the total observed fractionation. Note that Earth has the lowest D/H ratio among the solar system terrestrials{---VSWOW has D/H }$\sim$8 {times the solar abundance} \citep[c.f.][]{Hagemann:1970,Asplund:2009}. Measurements of large isotopic fractionations in the atmospheres of nearby exoplanets relative to their {host} stars would also likely represent departures from primordial compositions \citep{Molliere:2019}.

Atmospheric water vapor {observed in transmission }is suggestive but not definitive proof of the presence of an ocean (c.f. Earth, Venus; \citealt{Meadows:2018a,Lincowski:2018}). {Unlike reflectance spectroscopy, }transmission {spectroscopy} cannot {detect surface absorption or reflectance features}---additional observations would be needed to determine the likelihood of surface liquid water.

{Atmospheric escape is the only known mechanism capable of the extreme mass-dependent fractionation observed in D/H in our solar system \citep[see summary in][and references therein]{Molliere:2019}}.
Similarly, fractionation of oxygen (\ce{^18O/^16O}) during hydrodynamic or nonthermal escape of oxygen generated from photolysis of vaporized ocean water is another potential signature of {past} ocean loss. If the abiotic oxygen generated via ocean loss is not lost to space, but is instead dissolved into a magma ocean \citep{Schaefer:2016,Wordsworth:2018} or {oxidizes the} surface, then only a comparably small level of fractionation would occur.  Unlike atmospheric loss, adsorption by the surface {sequesters} heavier isotopes {slightly more than} lighter isotopes \citep[e.g.][]{Sharp:2017}, so would impart a small fractionation signature opposing that of escape.
Measuring the hydrogen and/or oxygen isotope fractionation for planets orbiting M~dwarfs could provide additional evidence of {past} extreme ocean loss and atmospheric escape in systems very different from our own.

Since \ce{O2} and its isotopologues are likely to be difficult to observe with \textit{JWST}, isotopologues of \ce{CO2} can be used as a more easily observed proxy for oxygen fractionation. Laboratory experiments \citep[e.g.][]{Shaheen:2007} and numerical modelling \citep{Liang:2007} of \ce{CO2} in the stratosphere of Earth have demonstrated rapid isotopic equilibrium (on the order of days) between \ce{CO2} and \ce{O2}, indicating that fractionation in \ce{CO2} can be efficiently induced if co-existing with heavily fractionated \ce{O2}.  
Since this process is UV-driven, it is likely to also occur efficiently {in the atmospheres of planets orbiting} M~dwarfs. 

Isotopic fractionation may therefore be a useful indicator for {past} ocean loss on terrestrial exoplanets and may help to observationally test the inner edge of the habitable zone. Here we assess how large isotopic fractionation could be observed spectroscopically in terrestrial exoplanet atmospheres with \textit{JWST}. We focus {primarily} on {the two TRAPPIST-1 planets most likely to produce the strongest transit signals, due to their large atmospheric scale heights and small semi-major axes: }TRAPPIST-1~b, which receives approximately twice the irradiation of Venus, and d, which lies between the conservative and optimistic IHZ {limits}. We also assess {the more observationally challenging TRAPPIST-1}~e, a habitable zone candidate. The TRAPPIST-1 system is scheduled to be observed with \textit{JWST} and is likely to produce a favorable signal that could be used to characterize evolved atmospheres \citep{Morley:2017,Lincowski:2018}. In \S\ref{sec:methods}, we summarize our models and methods, in \S\ref{sec:results} we show our results, in \S\ref{sec:discussion} we discuss the implications of our results for observations with \textit{JWST}, and in \S\ref{sec:conclusions} we summarize our findings.


\section{Methods} \label{sec:methods}

To produce {simulated} {transit transmission }spectra {\citep[see e.g.][and references therein]{Robinson:2017a} }with {increased} isotopic fractionation, we use a 1D line-by-line radiative transfer model and adjust the input line list isotopologue abundances. To assess simulated observations for \textit{JWST}, we use an instrument noise model.
We describe our models and inputs in the following subsections.

We adopt the following isotope geochemistry convention \citep[e.g.][]{Sharp:2017} for describing the isotopic fractionation of hydrogen (note this does not include the multiplier of 1000 typically used {in isotope geochemistry}, due to the extreme values we adopt here):
\begin{equation}
    \delta \ce{D} = \frac{(\ce{D/H})_\text{model}}{(\ce{D/H})_\text{VSMOW}}-1,
\end{equation}
and similarly, \ce{\delta^18O} {is the notation }for changes in \ce{^18O}/\ce{^16O}. ``VSMOW'' refers to the Vienna Standard Mean Ocean Water isotopic standard {\citep{Coplen:1995}}.

\subsection{Radiative Transfer}

To generate transmission spectra for analysis, we use the Spectral Mapping Atmospheric Radiative Transfer model \citep[SMART,][developed by D. Crisp]{Meadows:1996,Crisp:1997}. SMART is a 1D, line-by-line, multi-stream, multi-scattering model, which incorporates the Discrete Ordinate Radiative Transfer code \citep[DISORT,][]{Stamnes:1988,Stamnes:2000} to solve the radiative transfer equation. SMART has been shown to faithfully reproduce observed spectra of Mars \citep{Tinetti:2005}, Earth \citep{Robinson:2011}, and Venus \citep{Meadows:1996,Arney:2014}. SMART incorporates extinction from Rayleigh scattering and absorption from UV--visible electronic transitions, rotational-vibrational transitions, and collision-induced absorption (CIA). SMART can produce transit {transmission }spectra, including refraction \citep{Robinson:2017a}.

Rotational-vibrational absorption coefficients are calculated from line lists using our line-by-line absorption coefficients code \citep[LBLABC,][]{Crisp:1997}.  Here we have updated the partition functions \citep{HITRAN:2016-TIPS} and draw from the HITRAN2016 line list \citep{HITRAN:2016}, which are appropriate for the temperatures and pressures in these terrestrial atmospheres. We consider isotopic fractionation $\delta$D of 10--100, and $\delta$\ce{^18O} of 2--100, depending on the model atmosphere environment (see \S\ref{sec:justification}){, and adjust the abundances of the isotopologues accordingly (see \S\ref{sec:abundances})}.

We include collision-induced absorption for \ce{CO2}--\ce{CO2} \citep{Moore:1971,Kasting:1984,Gruszka:1997,Baranov:2004,Wordsworth:2010,Lee:2016}, \ce{N2}--\ce{N2} (\citealp{Schwieterman:2015b} based on \citealp{Lafferty:1996}), and \ce{O2}--\ce{O2} \citep{Greenblatt:1990,Hermans:1999,Mate:1999}.

\subsection{{Calculating Isotopologue Abundances}} \label{sec:abundances}

We determine the modified abundances for all isotopologues available for the molecules \ce{H2O}, \ce{CO2}, \ce{O3}, CO, and \ce{O2} in the HITRAN2016 line lists by calculating the abundances for each isotopologue given the specified isotopic fractionation. For oxygen we assume standard terrestrial linear mass fractionation, such that $\ce{\delta^17O}=0.5\ce{\delta^18O}$ \citep{Sharp:2017}, although around M~dwarfs this may depend on the details of atmospheric escape for a given planet. We assume that enhancement of doubly-fractionated molecules (e.g. \ce{D2O}) is stochastic, and therefore proportional to the production of the isotopic fractionation for each affected atom, which is generally a good assumption, especially at higher temperatures \citep{Eiler:2007}.

{To compute isotopologue abundances for each molecule, we numerically solve for the multipliers ($X$) for the isotopologues of a given molecule to adjust its VSMOW abundances given in HITRAN. These multipliers are not wholly independent, because we assume each substitution is proportional to the abundance. For example, substitution of hydrogen for deuterium in \ce{H2O}, H$_2^{17}$O, or H$_2^{18}$O  is equally likely per molecule, in proportion to the abundance of hydrogen in each molecule. The multiplier is squared for doubly-substituted isotopes, because it requires the probability that one atom is substituted \emph{and} the other is also substituted. A generic notation may be written as:
\begin{equation}
    X_{\ce{A_m B_n}} = \frac{(x_{\ce{A}} [\ce{A}])^m \cdot (x_{\ce{B}} [\ce{B}])^n}{[\ce{A}]^m \cdot [\ce{B}]^n},
\end{equation}
where the square brackets (e.g. [A]) are the notation for abundance and the individual multipliers are ($x_{\ce{A}}$) for a particular isotope A. This equation reduces to:
\begin{equation}
    X_{\ce{A_m B_n}} = x_{\ce{A}}^m x_{\ce{B}}^n.
\end{equation}
For our example of \ce{D2O} (also the example in box 1 of \citealt{Eiler:2007}), if $x_\text{D}$ is the multiplier for the enhancement in deuterium, the abundance multiplier of \ce{D2O} is:
\begin{equation}
    X_{\ce{D2O}} = \frac{(x_{\ce{D}} [\ce{D}])^2 \cdot  [\ce{^16O}]}{[\ce{D}]^2 \cdot[\ce{^16O}]} = x_{\ce{D}}^2.
\end{equation}
For a molecule with different isotopic substitutions, such as \ce{HD^17O}, the abundance adjustment would be:
\begin{equation}
    X_{\ce{HD^17O}} = x_{\ce{H}} x_{\ce{D}} (0.5 x_{\ce{^18O}}),
\end{equation}
where here we have set $x_{\ce{^17O}} = 0.5 x_{\ce{^18O}}$.
The multipliers $x$ are solved for simultaneously using a standard minimization code under the constraint that the total number of atoms of each isotope for a given family of molecules (e.g. \ce{H2O} and its isotopologues) satisfy the desired fractionation criterion (i.e. for \ce{\delta D}=100, that the ratio of abundances for deuterium across all isotopologues for a given molecule is 100 times greater than compared to VSMOW).}


We assume the line intensities (and absorption coefficients) of the adjusted isotopologues scale directly with their abundances, in accordance with the Boltzmann equation: 
\begin{equation}
    \frac{n_\text{iso}}{n_0} = \frac{g_\text{iso}}{g_0} e^{-(E_\text{iso}-E_0)/kT},
\end{equation}
where $n$ is the number of molecules in a given energy state (directly proportional to the absorption coefficient), $g$ represents the multiplicity of states ({here,} the abundance of molecules of a given isotopologue), $E$ is the energy of each state, $k$ is the Boltzmann constant, and $T$ is the temperature. 

\subsection{JWST Instrument Simulator}

We model the noise expected from our simulated spectral signals for various \textit{JWST} observing modes using the \textit{JWST} time-series spectroscopy simulator, PandExo\footnote{\url{https://natashabatalha.github.io/PandExo/}} \citep{Batalha:2017b}. PandExo uses Pandeia, the core of the Exposure Time Calculator of the Space Telescope Science Institute \footnote{\url{https://jwst.etc.stsci.edu/}} \citep{Pontoppidan:2016}. We consider only the optimal \textit{JWST} observing modes for these atmospheres as determined by \citet{Lustig:2019}, who conducted a comprehensive analysis of \textit{JWST} observing modes for the suite of atmospheres generated by \citet{Lincowski:2018}. As in \citet{Batalha:2018} and \citet{Lustig:2019}, for the NIRSpec Prism, we also consider an observing mode with a high-efficiency readout pattern by using a larger number of ``groups'' ($n_\text{groups}=6$) that allows saturation near the peak of the TRAPPIST-1 spectral energy distribution (SED), and therefore offers an improved duty cycle for the unsaturated spectral intervals. We do not impose a noise floor \citep[c.f.][]{Greene:2016}, as the on-orbit performance and systematic errors for \textit{JWST} are not currently known.

\subsection{Model Planetary Atmospheres and Stellar Inputs}

\iftwocol
    \input{table1.tex}
\fi

{To model post-ocean-loss atmospheres that have undergone isotopic fractionation, }we use nominal 10~bar atmospheres for TRAPPIST-1~b, d, and e from \citet{Lincowski:2018}. The choice of 10 bars is consistent with the findings of \ce{O2} sequestration by \citet{Wordsworth:2018}{, though other stable climate states with different compositions and higher or lower surface pressures and temperatures are also possible \citep[c.f.][]{Wolf:2017,Turbet:2018,Wunderlich:2019}.} TRAPPIST-1~b and d are the planets {that require the fewest transits} to observe {molecular absorption features, due to larger expected scale heights resulting from higher atmospheric temperatures, and a low surface gravity for d} \citep{Lincowski:2018,Lustig:2019}. Planet d also sits between the {current conservative and recent Venus }estimates for the inner edge of the habitable zone{, and so could help constrain the actual inner  limit of the HZ for the TRAPPIST-1 system}. TRAPPIST-1~e is firmly in the conservative HZ, and is perhaps most likely to be temperate {\citep{Wolf:2017,Turbet:2018,Lincowski:2018}}. {Though} some or all of the TRAPPIST-1 planets could have much larger water abundances {because their densities are generally lower than the density of Earth, here we assume initially terrestrial bulk compositions, as their $3\sigma$ error bars also encompass the density of Earth} \citep{Grimm:2018}. {For this isotopic fractionation detection study, we have assumed that all three of these planets, including TRAPPIST-1~e, have lost their surface water due to an early runaway greenhouse phase and atmospheric escape during the superluminous pre-main-sequence of the host star, and so are not habitable \citep{Lincowski:2018}.}

{For all three planets, we modeled} \ce{O2}-dominated {atmospheres, both} desiccated and outgassing, and clear-sky Venus atmospheres. A cloudy{/hazy} Venus case is not included for the detectability studies considered here, as haze opacity {(due to Mie scattering at $<2.5$\um{} and absorption by \ce{H2SO4} at $>2.5$\um{}, \citealt{Palmer:1975,Pollack:1993,Ehrenreich:2012,Lincowski:2018}) likely} precludes detection of the isotopologue bands. {Furthermore, due to high temperatures, \ce{H2SO4} will not likely condense in the atmosphere of a Venus-like TRAPPIST-1~b \citep{Lincowski:2018}}.
We model a second set of spectra for the \ce{O2}-dominated atmospheres that reduce the water abundances for b and d by a factor of 100 and 10 respectively, to simulate lower outgassing rates (and a drier stratospheric water abundance of $\sim$1~ppm) than those assumed in \citet{Lincowski:2018}. 

\iftwocol
\else
    \input{table1.tex}
\fi

The model atmospheres are detailed in \citet{Lincowski:2018}{ and listed in Table~\ref{table:environments}}. Briefly, they contain 58--66 levels from the surface to 0.01~Pa. The \ce{O2} desiccated atmosphere is 95\% \ce{O2}, 0.5\% \ce{CO2}, and 4.5\% \ce{N2} in photochemical--kinetic equilibrium with the primary photolytic products CO and \ce{O3} \citep[see][their Figure 4]{Lincowski:2018}. The \ce{O2} outgassing atmosphere is 95\% \ce{O2} and 4.5--5\% \ce{N2}, with Earth-like volcanic outgassing fluxes at the surface for \ce{H2O}, \ce{CO2}, \ce{SO2}, and other molecules not detectable in these spectra \citep[see][their Figures 4 and 14]{Lincowski:2018}. {Note that the \ce{O2} atmospheres with reduced water abundance are not climatically or photochemically self-consistent.} The Venus-like atmosphere is 96.5\% \ce{CO2}, 3.5\% \ce{N2}, with trace amounts of \ce{H2O}, \ce{SO2}, and others fixed to values at the surface consistent with Venusian abundances at 10~bar \citep[][ their figure 6]{Lincowski:2018}.

These atmospheres were generated using a 1D line-by-line, multi-stream, multi-scattering radiative-convective equilibrium climate model coupled to a photochemical--kinetics model.
More information about the climate-photochemical and spectral modeling of these model atmospheres can be found in \citet{Lincowski:2018}, particularly their \S2 (Model and Input Descriptions) and \S3 (Results).
As in \citet{Lincowski:2018}, we use the updated planetary masses and radii from \citet{Grimm:2018}, and the semi-major axes and stellar radius from \citet{Gillon:2017} to compute transit spectra.

\subsection{Isotopic Fractionation} \label{sec:justification}

\iftwocol
    \input{table2.tex}
\fi

Since isotopologues have not been considered in M~dwarf terrestrial planetary atmospheric escape modeling \citep{Lammer:2007,Luger:2015a,Luger:2015b,Schaefer:2016,Ribas:2016,Airapetian:2017,Bolmont:2017,Dong:2017,Dong:2018,Wordsworth:2018,Lincowski:2018,Egan:2019}, we use Venus observations to constrain plausible fractionation values for our spectral modeling.
While the Venus literature is mostly in agreement that hydrodynamic loss was responsible for the primordial loss of water \citep{Hunten:1982,Kasting:1983,Kasting:1988,Chassefiere:1996,Gillmann:2009,Chassefiere:2012,Bullock:2013,Lichtenegger:2016,Lammer:2018} {and may have been responsible for fractionation of D/H \citep[c.f. equation (17)][]{Hunten:1987}}, the specific mechanism(s) that caused the current D/H fractionation are uncertain (e.g. \citealt{Kasting:1983,Kasting:1988,Grinspoon:1993,Gillmann:2009,Collinson:2016,Lichtenegger:2016} and reviews by \citealt{Chassefiere:2012,Bullock:2013,Lammer:2018}).
These processes generally cause some degree of fractionation and favor escape of the lighter elements, either due to the larger escape energy required by heavier species or due to the diffusive stratification of the homosphere and resultant higher abundances of lighter elements near the exobase \citep[i.e. Rayleigh fractionation,][]{Rayleigh:1896,Hunten:1982,Sharp:2017}.

Higher fractionation values compared to Venus may be possible due to the more extreme stellar radiation environment experienced by M~dwarf planets. M~dwarfs have a much longer superluminous pre-main-sequence phase \citep{Baraffe:2015} and they emit comparatively more XUV and FUV flux than our Sun \citep[e.g.][]{Ribas:2016}, which enhances atmospheric escape over time through stronger ionospheric heating from XUV absorption, and stronger photolysis of water vapor by FUV absorption, a key limiting factor defining the diffusion-limited escape flux \citep[e.g.][]{Luger:2015b}.
Terrestrial planets in and around the habitable zones of M~dwarf stars could lose hundreds of bars of oxygen to nonthermal escape processes, such as CME- or solar-wind-driven ion pick-up \citep[c.f.][]{Kulikov:2006,Lammer:2007,Gillmann:2009} and polar winds \citep[c.f.][]{Collinson:2016,Airapetian:2017}, both of which may be aggravated for planets without strong (i.e. Earth-like) magnetic fields. Although these studies have not considered isotopes or isotopic fractionation, the large loss potential for oxygen and other heavier species may plausibly result in severe isotopic fractionation. 

Without detailed calculations of the possible isotopic fractionation in the atmospheres of planets around M~dwarf stars, we assume a range of values for \ce{\delta D} and \ce{\delta^18O} both consistent with and more severe than Venus.
For the environments with water vapor, we {conservatively} simulate \ce{\delta D} up to 100 times VSMOW, consistent with Venus ($\sim$120--140). {For cases where \ce{\delta D}=100}, we also simulate \ce{\delta^18O} up to 10 times VSMOW (note Venus exhibits no oxygen fractionation). In our most severe case for atmosphere and ocean loss, we simulate \ce{\delta^18O} up to 100 for the desiccated, \ce{O2}-dominated environment. For this extreme case, we assume that early complete ocean and hydrogen loss was followed by continued escape of oxygen. This severe value may not be possible, but it is useful to calculate a range of values in these spectral experiments to demonstrate the thresholds for detection. We include lesser fractionation values as appropriate for each case.

\section{Results} \label{sec:results}

We present noiseless simulated {transit transmission }spectra {at 1~cm$^{-1}$ resolution} demonstrating the signal present due to different levels of extreme isotopic fractionation for $\delta$D up to 100 VSMOW (similar to Venus) and \ce{\delta^18O} up to 100 VSMOW. {These spectra are presented as ``relative transit depth'', {given as \citep[c.f.][]{Winn:2010,Lincowski:2018}:}}
\begin{equation} \label{eq:dF_F}
   {\frac{dF_a}{F} = \frac{2 R_p R_a}{R_*^2} + \left( \frac{R_a}{R_*} \right)^2,}
\end{equation}
{where $F$ is the stellar flux, $dF_a$ is the difference in stellar flux due to occultation by the atmosphere of the transiting planet, $R_p$ is the planet solid-body radius, $R_a$ is the atmospheric height from the surface of the planet, and $R_*$ is the radius of the star. In this work, the relative transit signals discussed are generally relative to the transit signal calculated with the nominal VSMOW abundances.} {The spectra and data are available online using the VPL Spectral Explorer\footnote{\label{footnote:specex}\url{http://depts.washington.edu/naivpl/content/vpl-spectral-explorer}}, or upon request.}
We assess the detectability of {the isotopically-enhanced features of }these spectra propagated through the PandExo \textit{JWST} instrument noise simulator {\citep{Batalha:2017b}}.

\subsection{Isotopologue Abundances}

\iftwocol
\else
    \input{table2.tex}
\fi

We solve for the isotopologue abundances of the model atmospheres we adopt from \citet{Lincowski:2018} numerically {as described in \S\ref{sec:abundances}}. The calculated abundances for all fractionation values for the primary detectable species, \ce{H2O} and \ce{CO2}, are listed in Table~\ref{table:abundances}, which are listed in columns for each iteration of \ce{\delta{D}} and \ce{\delta^18O}.

\subsection{Simulated Spectra}

\begin{figure*} 
  \centering
  \iftwocol
    \includegraphics[width = \textwidth]{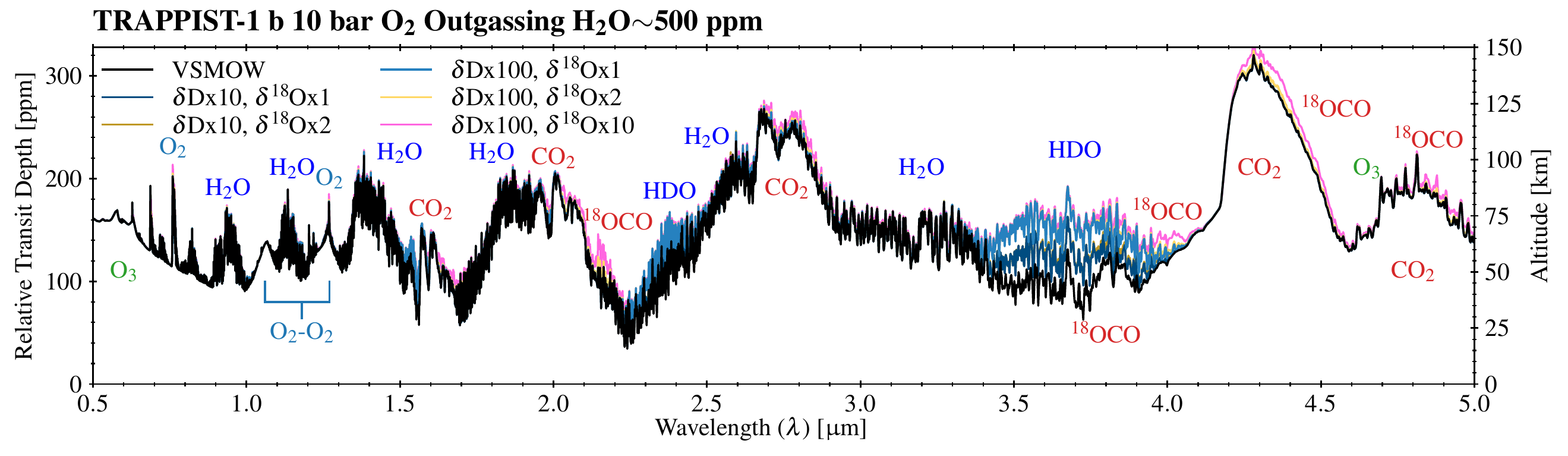}
    \includegraphics[width = \textwidth]{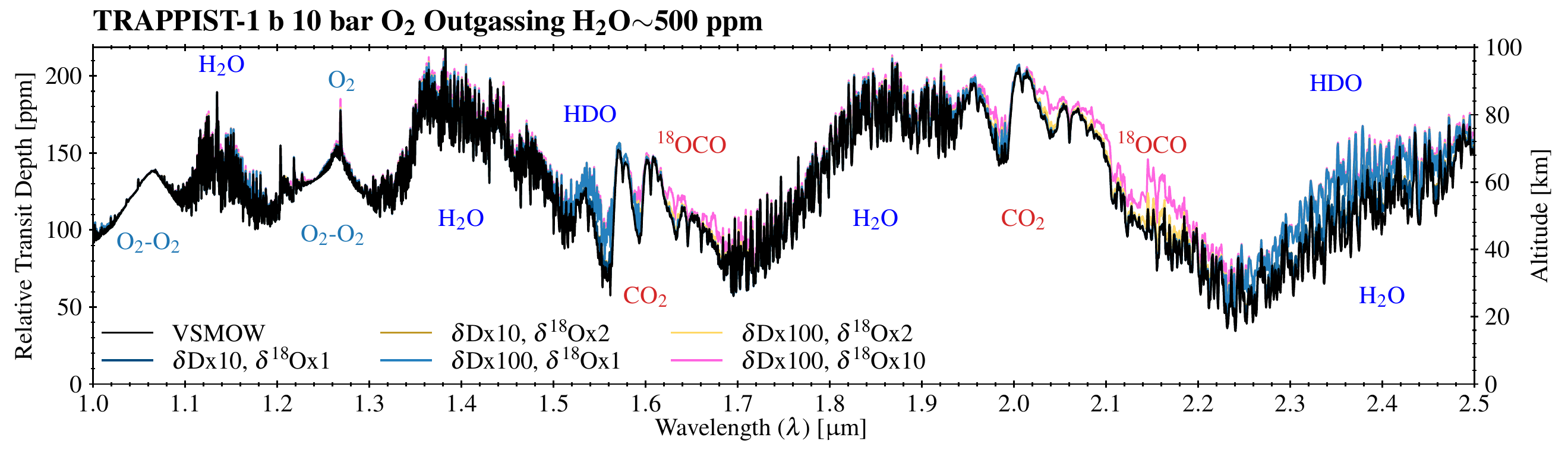}
    \includegraphics[width = \textwidth]{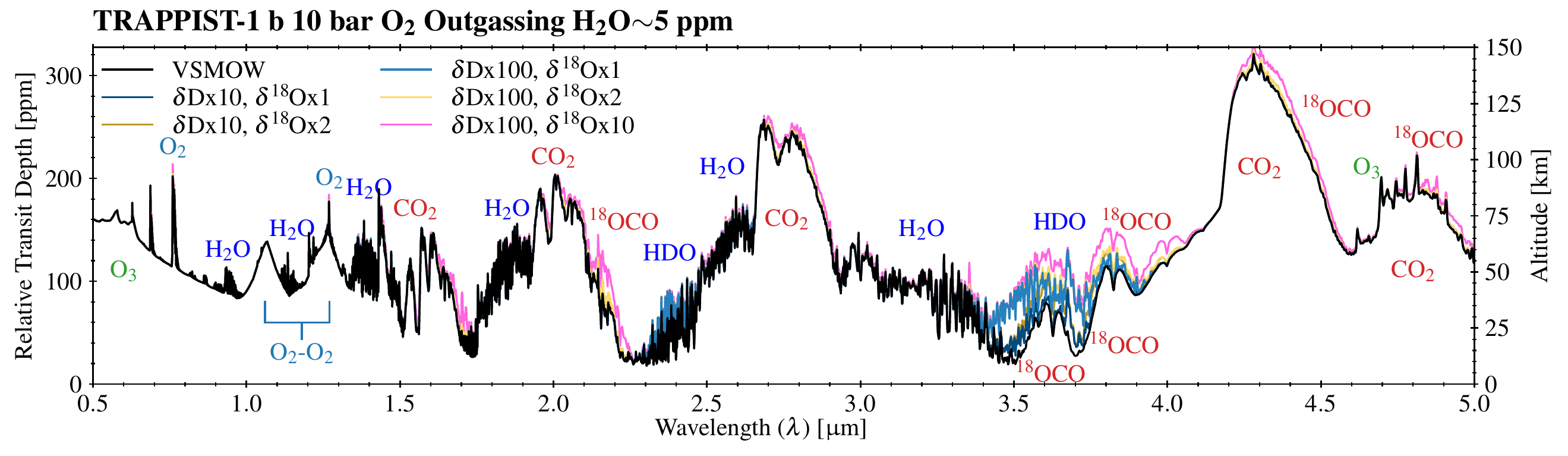}
  \else
    \includegraphics[width = \textwidth]{t1b_10bar_o2}
    \includegraphics[width = \textwidth]{t1b_10bar_o2_10_25}
    \includegraphics[width = \textwidth]{t1b_10bar_o2_ppm} 
  \fi
  \caption{\textbf{\textsc{Simulated TRAPPIST-1~b atmospheres exhibit isotopologue features that may be detectable with \textit{JWST}.}} Simulated transmission spectra for {the \ce{O2} outgassing} environments for TRAPPIST-1~b{, which contain} $\sim$500~ppm or $\sim$5~ppm stratospheric \ce{H2O}, comparing up to \ce{\delta{D}=100} and \ce{\delta^18O=10}. {These} atmospheres exhibit transmission signals at multiple wavelengths up to 79~ppm, primarily from HDO. {High HDO abundances partially obscure signals from \OCO{}{ (notated as \ce{^18OCO} in the plots)}. The center panel shows the range of highest signal from the star, 1.0--2.5~\um{}, which contains weak isotopologue bands that may contribute significantly to the detection of these species. }  \label{fig:noiseless}
  }
\end{figure*}

\begin{figure*} 
  \centering
  \iftwocol
    \includegraphics[width = \textwidth]{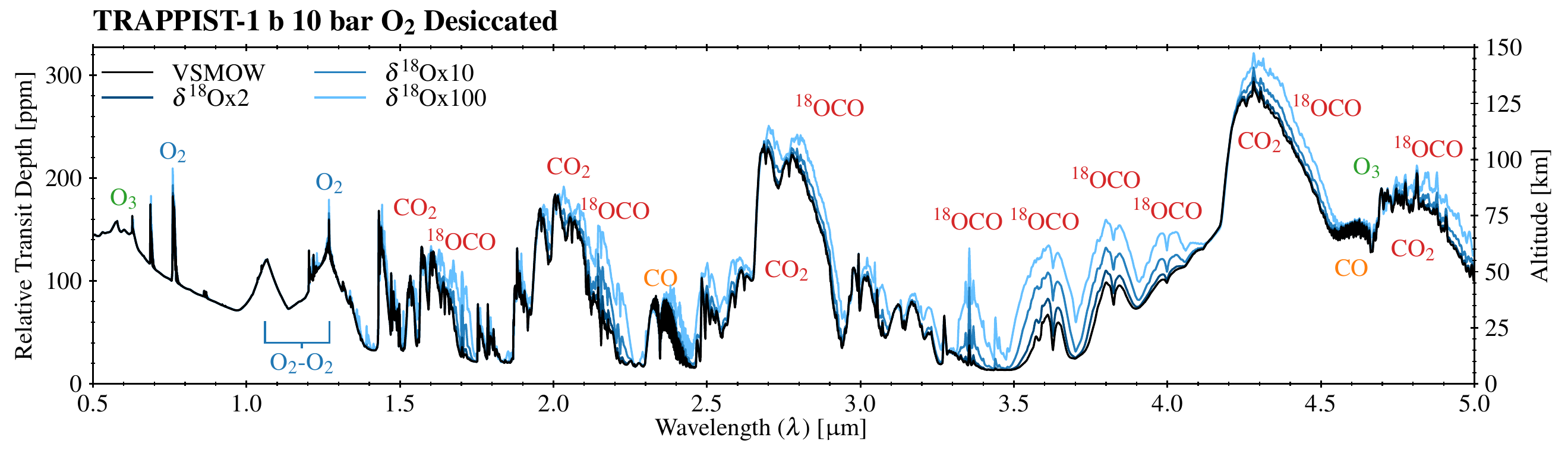}
    \includegraphics[width = \textwidth]{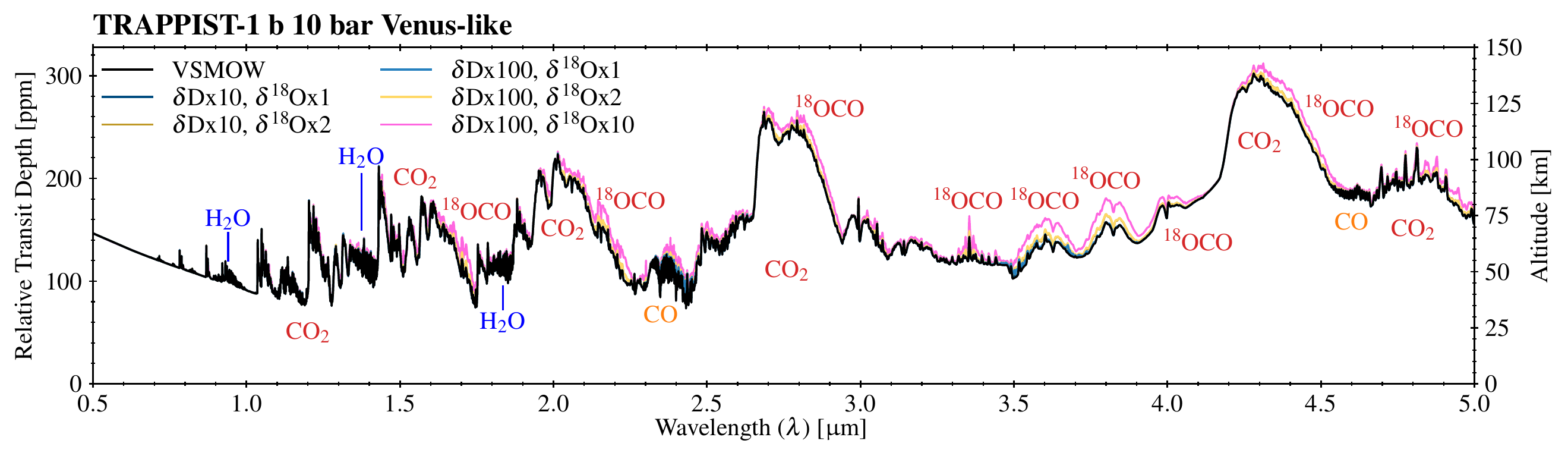}
  \else
    \includegraphics[width = \textwidth]{t1b_10bar_o2_dry}
    \includegraphics[width = \textwidth]{t1b_10bar_co2}
  \fi
  \caption{\textbf{\textsc{Simulated TRAPPIST-1~b atmospheres exhibit isotopologue features that may be detectable with \textit{JWST}.}} {\textit{Upper panel:}} simulated transmission spectra for a 10~bar desiccated \ce{O2}-dominated atmosphere, demonstrating \ce{\delta^18O} in \ce{CO2} up to 100. \textit{Lower panel}: {simulated transmission spectra for a }10~bar Venus-like (\ce{CO2}-dominated) atmosphere comparing up to \ce{\delta{D}=100} and \ce{\delta^18O=10}. This case does not have clouds. The desiccated \ce{O2}-dominated atmospheres exhibit up to 94~ppm transit signals at multiple wavelengths, primarily from \OCO{}{ (notated as \ce{^18OCO})}. The Venus-like atmosphere exhibits weak isotopologue fractionation signals because the spectrum is dominated by strong \ce{CO2} absorption.  \label{fig:noiseless_2}
  }
\end{figure*}

\begin{figure*} 
  \centering
  \includegraphics[width = \textwidth]{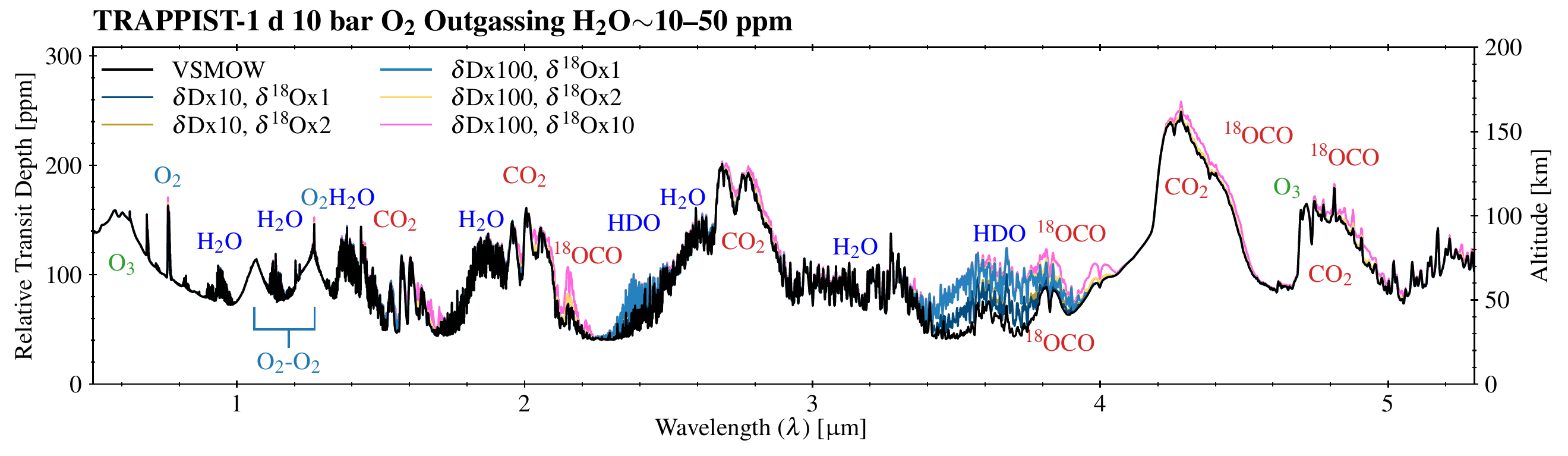}
  \includegraphics[width = \textwidth]{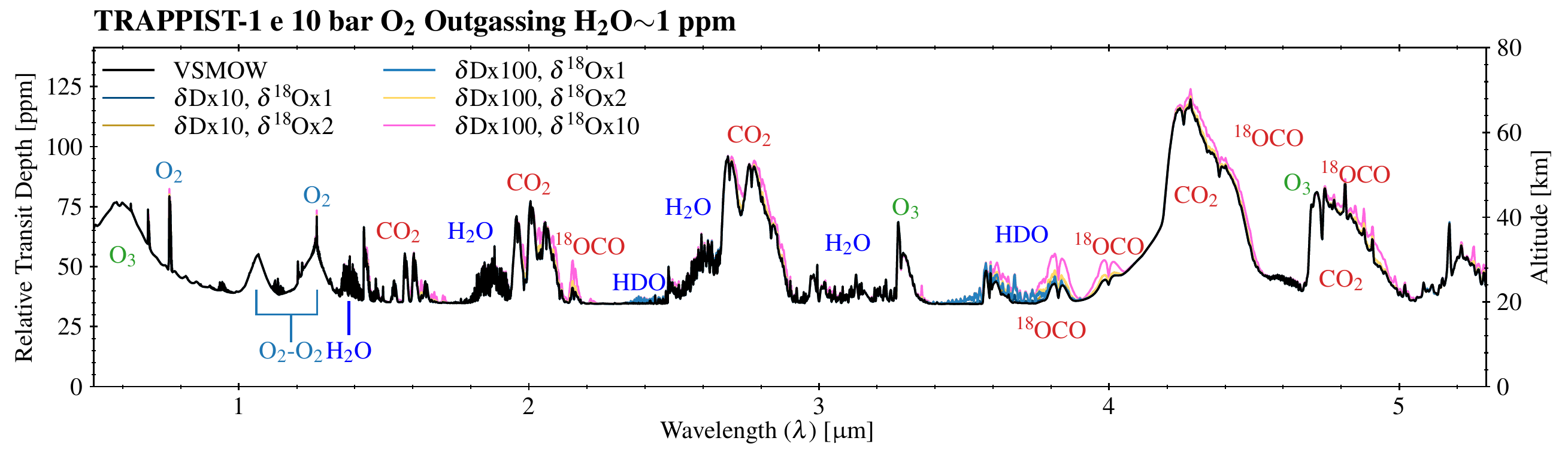}
  \caption{\textbf{\textsc{TRAPPIST-1~d and e \ce{O2}-dominated atmospheres with geological outgassing may exhibit detectable isotopologue features in transit transmission spectra.}} Simulated transmission spectra for TRAPPIST-1~d (upper panel) and e (lower panel) compare up to \ce{\delta{D}=100} and \ce{\delta^18O=10}. These spectra are similar to the TRAPPIST-1~b case. Planets d and e are farther away from the star and exhibit more refraction of stellar photons in transit, which reduces the sizes of the absorption features. For d, transit transmission signals peak at 55~ppm for \ce{\delta{D}=100} and 36~ppm for \ce{\delta^18O=10}, with combined features peaking at 57~ppm. For e, transit transmission signals peak at 11~ppm for \ce{\delta{D}=100} and 12~ppm for \ce{\delta^18O=10}.
  \label{fig:t1d}}
\end{figure*}

We assessed isotopic fractions of $\delta$D up to 100 times VSMOW for two outgassing, \ce{O2}-dominated atmospheres, and a (clear-sky) Venus-like, \ce{CO2}-dominated atmosphere, for TRAPPIST-1~b, d, and e. We assessed \ce{\delta^18O} up to 10 for these atmospheres, and up to 100 VSMOW for a completely desiccated \ce{O2}-dominated atmosphere. Justification for these values was discussed in \S\ref{sec:justification}{ and these environments are listed in Table~\ref{table:environments}}.

Noiseless simulated spectra are shown in Figures~\ref{fig:noiseless}--\ref{fig:noiseless_2} for TRAPPIST-1~b. The spectral region around 3--4~\um{} is rich for observing $\delta$D enhancements in HDO and \ce{\delta^18O} enhancements in \OCO{}. TRAPPIST-1~b exhibits transit transmission signal strengths up to {79}~ppm for \ce{\delta{D}} in the reduced \ce{H2O}, \ce{O2}-dominated atmosphere with outgassing and 94~ppm for \ce{\delta^18O} in the desiccated \ce{O2}-dominated atmosphere, compared to nominal spectra with VSMOW abundances. For comparison with b, in Figure~\ref{fig:t1d} we show simulated spectra for{ TRAPPIST-1~d and e for} the \ce{O2}-dominated atmospheres with outgassing. {The atmospheres of both planets} have smaller absorption features due to lower {water vapor abundances and }temperatures, and {due to }the refraction of stellar photons. TRAPPIST-1~e has very small isotopologue features here compared to VSMOW abundances, but this {simulated, uninhabitable} environment has a temperate climate and stratospheric water abundance similar to a potentially habitable planet, which makes it an important comparison case.

The transit signals of isotopic fractionation in our individual atmospheres vary considerably and depend on abundances of water vapor and \ce{CO2} (Figure~\ref{fig:noiseless}). In an \ce{O2}-dominated atmosphere with some water vapor still present due to outgassing, Venus-like fractionation of $\delta$D could be observable. With $\sim$500~ppm stratospheric \ce{H2O} (Figure~\ref{fig:noiseless}, upper panel), the modeled TRAPPIST-1~b atmosphere may exhibit a transit signal up to {74}~ppm in the broad 3.7~\um{} HDO band. There are also weaker features due to HDO at 1.5 and 2.4~\um{} if HDO is sufficiently abundant. If fractionation in oxygen was also present, the \OCO{} features between 3.0--4.1~\um{} would overlap with the 3.7~\um{} HDO band, though the band widths and shapes are different. The climate-photochemistry models of \citet{Lincowski:2018} calculated lower water abundances in the stratosphere of this environment for TRAPPIST-1~d (10--50~ppm) and e ($\sim$1~ppm), and the transit signal for enhanced HDO is similarly smaller (55 and 11~ppm, respectively){. These transit signals are also} reduced due to refraction for both planets and a much smaller scale height for e. With reduced stratospheric water vapor (to $\sim$1--5~ppm each for b and d), compared to VSMOW the spectra exhibited isotopologue transit signals of {79} and {29}~ppm for b and d, respectively.

For a more severe case in a desiccated (water-free) \ce{O2}-dominated environment (Figure~{\ref{fig:noiseless_2}, upper} panel), enhancements to \ce{\delta^18O} in \OCO{} are evident throughout the NIR (strongest at 1.7, 2.2, and 3--4~\um{}{; labeled as \ce{^18OCO}}), exhibiting transit signals compared to VSMOW of up to 94~ppm for b and {72}~ppm for d. With a signal of {29}~ppm, this was the only TRAPPIST-1~e atmosphere to exhibit a transit signal {near} 30~ppm compared to VSMOW {\citep[i.e. the putative noise floor for NIRSpec,][]{Greene:2016}}. There are also strong \OCO{} isotopologue bands in these atmospheres at 6.0, 7.3, and 7.9~\um{} (not shown), though this spectral region is less favorable for \textit{JWST} observations \citep{Morley:2017,Lustig:2019}.

For all three modeled planets (b, d, and e), Venus-like atmospheres are thoroughly dominated by \ce{CO2}, {with minimal differences in spectral features due primarily to \OCO{}} (less than 30~ppm) in the transit signal compared to VSMOW{ (see e.g. Figure~\ref{fig:noiseless_2}, lower panel)}. These differences were at the same wavelengths exhibited by the \ce{O2} atmospheres. With \ce{CO2} absorption saturating the NIR spectrum, {HDO was not detectable }in our Venus-like cases. 

{While HDO and \OCO{} were the primary detectable isotopologues, many other bands were included in our spectral models, but were generally not present in the simulated transit transmission spectra, including the \ce{O2} A-band. No isotopologues containing \ce{^17O} were distinctly present, due to its lower abundance compared to \ce{^18O}. Isotopologue absorption by \ce{^12C^18O} at 2.4~\um{} is distinguishable in the completely desiccated atmosphere (see Figure~\ref{fig:noiseless}), but not likely individually discernible with \textit{JWST}.  }

\subsection{Detectability Assessment}

\begin{figure*} 
  \centering
  \includegraphics[width = \textwidth]{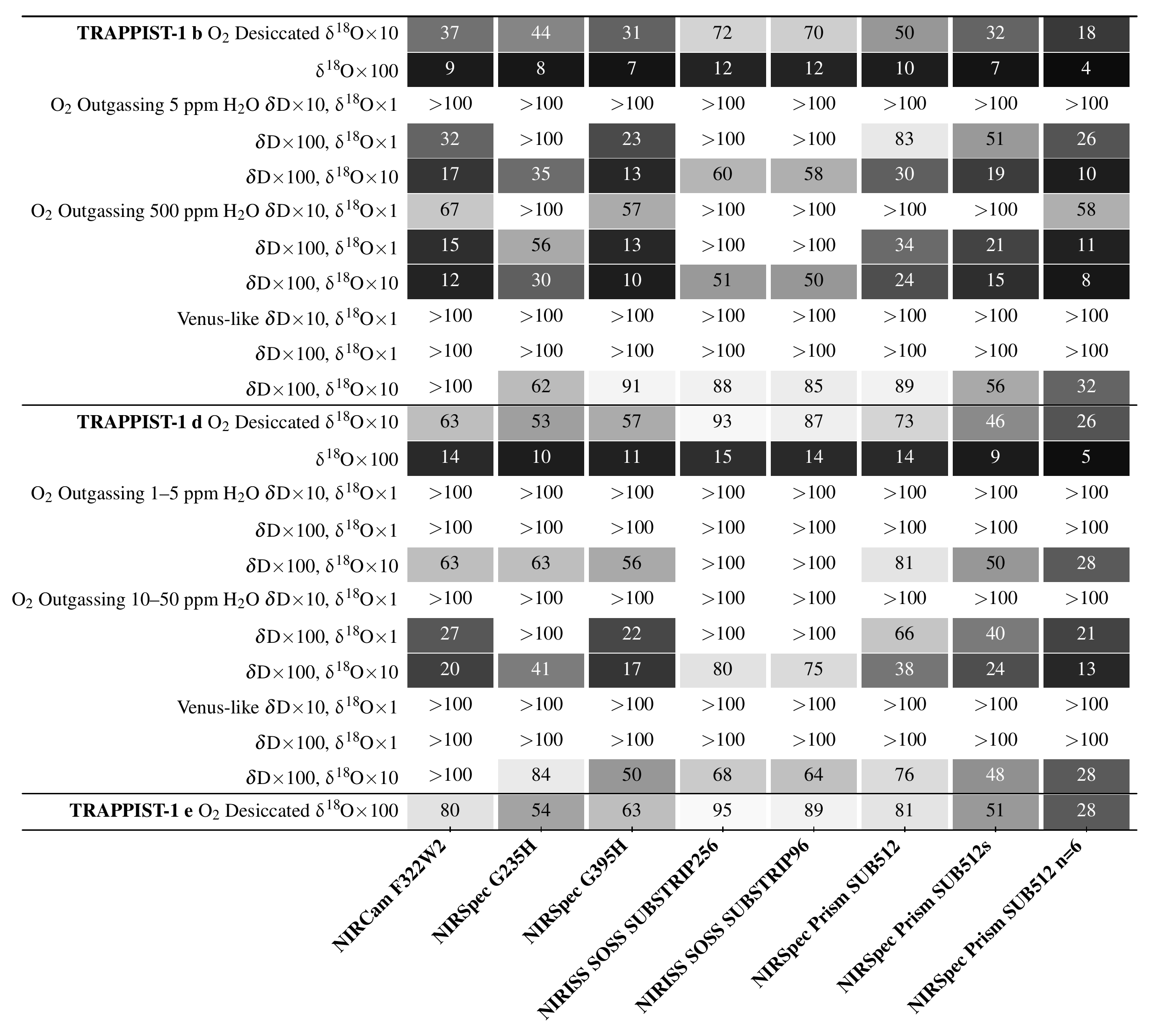}
  \caption{\textbf{\textsc{Transits to distinguish isotopologue features {\textsc{compared to}} VSMOW {\textsc{abundances}} at \SN{}~=~5 with a selection of \textit{JWST} instruments and for the modeled atmospheres.}} NIRSpec Prism and G395H Grism, along with NIRCam with F322W2 filter, are best suited to detect HDO and \OCO. All modeled TRAPPIST-1~e spectra, except for the case shown, required greater than 100 transits to distinguish isotopologue features compared to VSMOW {abundances} at \SN{}=5. {The cells are colored darker for fewer transits and lighter for more transits required.} 
  \label{fig:detectability}}
\end{figure*}

\begin{figure*} 
  \centering
  \iftwocol
    \includegraphics[width = \textwidth]{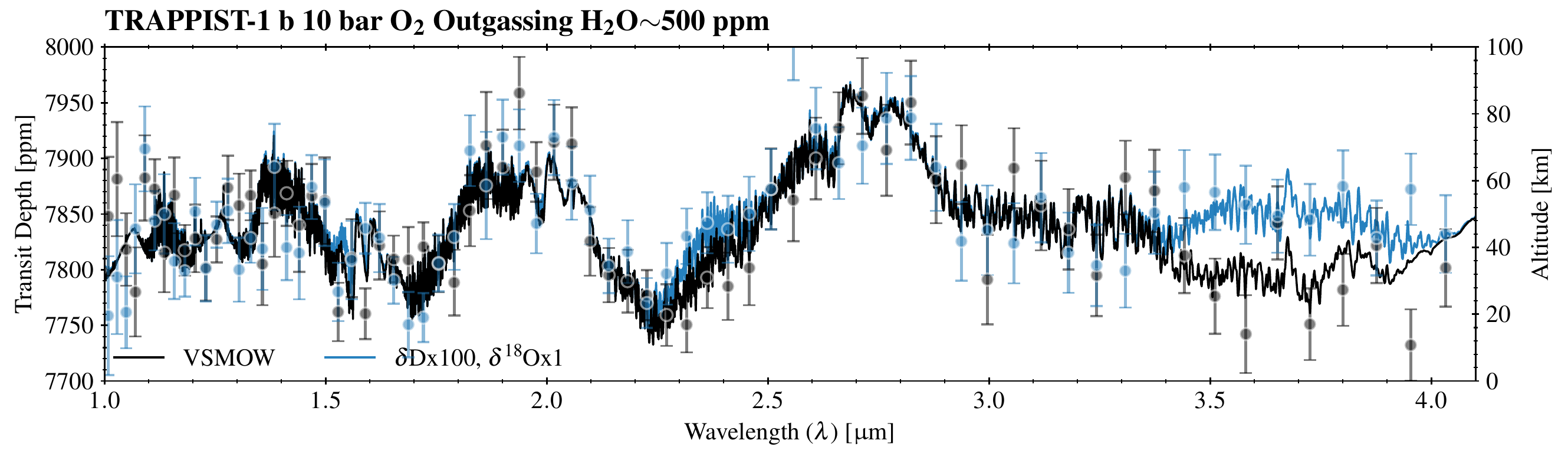}
  \else
    \includegraphics[width = \textwidth]{t1b_o2_noise}
  \fi
  \caption{{\textbf{\textsc{Isotopologue bands may be detected in a TRAPPIST-1~b clear-sky post-ocean-loss \ce{O2}-dominated atmosphere with \textit{JWST}.}} Simulated transmission spectra for the \ce{O2} outgassing environment for TRAPPIST-1~b with $\sim$500~ppm stratospheric \ce{H2O}, comparing \ce{\delta{D}=100} (blue) vs VSMOW (black). These spectra were processed through the PandExo \textit{JWST} instrument simulator \citep{Batalha:2017b} and are presented with error bars at the native resolution for NIRSpec Prism (nominal resolving power $R\sim100$), using $n_\text{groups}=6$ \citep{Batalha:2018}, with 11 transits co-added. The difference between these spectra is detectable at \SN{}=5. The broad HDO band at 3.7~\um{} provides significant contribution to the overall detection. The \ce{O2}-\ce{O2} bands at 1.06 and 1.27~\um{} can also be seen here.}  \label{fig:noise}
  }
\end{figure*}


\citet{Lustig:2019} conducted a comprehensive detectability study for \textit{JWST} instruments using the suite of model spectra from \citet{Lincowski:2018} and identified several useful instrument modes for these atmospheres{. These optimal instrument modes were used} here{, and are listed along the bottom of Figure~\ref{fig:detectability}}, {which provides} a summary of {the number of transits required to attain an expected signal-to-noise (\SN{}) of five compared to the transit spectra with nominal VSMOW abundances for each model atmosphere. The number of transits required for a different \SN{} can be calculated by: }

\begin{equation}
    {N' = N \left( \frac{ \left<\text{S/N}\right>' } { \left<\text{S/N}\right> } \right)^2,}
\end{equation}
where $N$ is the number of transits (here at \SN{}=5) and $N'$ is the number of transits to attain \SN{}$'$.

Except for our {\ce{CO2}-dominated, }Venus-like environments, {Venus-like} levels of fractionation {of D/H} consistent with {past} ocean loss for each TRAPPIST-1~b and d atmosphere may be detectable with \textit{JWST} {(see e.g. Figure~\ref{fig:noise})}. The completely desiccated, \ce{O2}-dominated atmospheres {modeled for TRAPPIST-1~b and d have \OCO{} bands that} are {also} accessible to \textit{JWST}. {Assuming volcanically-outgassed species in the \ce{O2}-dominated atmospheres allows detection of HDO bands. Fractionation in oxygen, in addition to hydrogen, increases the detectability of isotopologue bands due to complementary signal increases in both HDO and \OCO{} bands. As in \citet{Lustig:2019}, {we find that }NIRSpec Prism SUB512 {(512x512 subarray) with a partial saturation strategy (allowing }partial saturation at the SED peak to {provide} higher signal at other wavelengths), with {six groups per integration} {\citep{Batalha:2018}}, is generally the ideal instrument/mode for these detections. NIRSpec G395H grism is nearly as good, or in some cases better, due to coverage of the 3--4~\um{} region containing broad HDO and \OCO{} bands. Similarly, due to its wavelength coverage, NIRCam F322W2 is also an acceptable instrument for these detections.}

With lower levels of stratospheric water vapor similar to Venus and Earth in an \ce{O2}-dominated atmosphere (here, 1--5~ppm \ce{H2O}), {a similar number of transits are required for TRAPPIST-1~b }compared to the nominal outgassing atmospheres of \citet{Lincowski:2018}. This is significantly worse for {TRAPPIST-1}~d{, which may be in part due to the higher refraction altitude exhibited by d.}

The Venus-like atmosphere with Venus-like isotopic fractionation is likely very difficult to detect with \textit{JWST}{, even without aerosols}. This is due to the small scale height and high opacity of a \ce{CO2}-dominated atmosphere, which confines the transmission to the upper stratosphere and strongly reduces the signal from \ce{H2O} and HDO. To detect {a fractionation signal}, such a Venus-like planet would also have to exhibit substantial fractionation in oxygen (unlike Venus itself), which could enhance the transit depths of the \OCO{} bands. In a cooler atmosphere such as that of TRAPPIST-1~d, the \ce{CO2} bands are not as {broadened}, and the occupancy of the lower energy states of isotopologue bands is higher \citep{Molliere:2019}, so detecting \OCO{} in a Venus-like (albeit clear-sky) atmosphere of d would require {fewer transits than b}.


For TRAPPIST-1~e, detecting isotopic fractionation in \ce{H2O} would be difficult, requiring more than 100 transits for all fractionation values, atmospheric environments, and instrument modes considered here. Fractionation in \ce{CO2} oxygen isotopes in the desiccated \ce{O2}-dominated atmosphere may be possible{, if such extreme fractionation is possible}. {The HDO and \OCO{} features in the model atmospheres for TRAPPIST-1~e are substantially masked by atmospheric refraction. The cooler temperatures and higher surface gravity compared to TRAPPIST-1~b and d result in small scale heights, which conspire to produce shallow transmission depths.}

\section{Discussion} \label{sec:discussion}

We have shown that an enhancement of 100 in either $\delta$D (via water vapor) or \ce{\delta^18O} (via \ce{CO2}){, which would likely be} due to extreme ocean or atmospheric loss, may be detectable with \textit{JWST} for the inner planets of the TRAPPIST-1 system. Simulated spectra for our \ce{O2}-dominated atmospheres exhibit transit signals up to {79}~ppm and 94~ppm for $\delta$D and \ce{\delta^18O} respectively for TRAPPIST-1~b (55 and {72}~ppm respectively for d; 11 and {29}~ppm respectively for e). With optimal use of NIRSpec Prism \citep{Batalha:2018}, these signals could be detected at \SN{}=5 in {eleven} and four transits respectively for b ({25} and five for d). In the {desiccated, \ce{O2}-dominated} case with \ce{\delta D}~=~100 and \ce{\delta ^18O}~=~10, the isotopic enhancement can be distinguished from the nominal Earth VSMOW abundances in as few as {eight} transits for b ({13} for d). {This is only a few more transits than} the two transits required with NIRSpec Prism {SUB512 $n=6$ }to {detect an atmosphere by ruling} out a featureless spectrum at \SN{}~=~5 for b or d \citep{Lustig:2019}. It is more difficult to observe a fractionation signature in the atmosphere of a habitable zone planet like TRAPPIST-1~e; only extreme fractionation of oxygen is possibly detectable (in {28} transits), due to a smaller atmospheric scale height, lower levels of stratospheric water vapor, and the refraction of stellar rays through the atmosphere during transit \citep{Betremieux:2013,Betremieux:2014,Misra:2014,Robinson:2017a}.

{The strongest transit signal is due to rotational-vibrational bands in the NIR, 1--5~\um{}, consistent with \textit{JWST} NIRSpec. Transit observations are dependent on the stellar photons for signal, and this is the spectral region with the most stellar photons from M~dwarf stars. These rotational-vibrational transitions are subject to mass-dependent spectral shifts for the bands between 1--3~\um{}, because such transitions are inversely proportional to the reduced mass of the molecule. The asymmetric molecules such as HDO and \OCO{} provide new degrees of freedom due to breaking molecule symmetries compared to \ce{H2O} and \ce{CO2}, which produce new absorption bands between 3--8~\um{}.}

{The severe fractionation considered in this work would not likely be caused by any other known fractionation mechanism, including the composition of the host star \citep[e.g.][and references therein]{Molliere:2019}.} Interpreting observed isotopic abundances in the atmospheres of an exoplanet must be considered in relation to the host star{. However, nearby stars differ in D/H by less than a factor of $\sim2$ compared to the Sun \citep[c.f.][]{Linsky:2006,Asplund:2009}.} Fractionation of isotopic abundances compared to the host star value {would} indicate that a planet evolved from the primordial conditions, which could include a small effect due to formation. However, a large fractionation compared to the host star would likely indicate an evolved atmosphere, though at this time we cannot say which mechanism may be responsible.

The ability to observe isotopic fractionation in \ce{H2O} or \ce{CO2} {may} help to more robustly identify worlds without surface oceans.  Transmission {spectroscopy} cannot {directly} probe the surface {composition} of a planet, and water vapor observed in transmission is suggestive, but is not conclusive proof that an ocean is present. 
{Current planetary hydrogen loss (though also not conclusive evidence of a surface ocean) could be identified via UV measurements of Lyman-$\alpha$ \citep[e.g.][]{Jura:2004}.}
The detection of \ce{O2}-\ce{O2} collision-induced absorption is indicative of a large inventory of oxygen that is most likely produced during the loss of oceans of water \citep{Luger:2015b, Schwieterman:2016}{. A large \ce{O2} inventory }does not preclude the possibility that a significant amount of surface water remains, especially for more volatile rich worlds, which may include the TRAPPIST-1 planets. However, the identification of isotopic fractionation in \ce{H2O} and/or \ce{CO2} in conjunction with the detection of \ce{O2}-\ce{O2} would be strong evidence that the planet lost the bulk of its available water inventory, and that it is unlikely that surface water remains.  This is because an extant inventory of surface water would dilute the isotopic fractionation signal.  

Isotopic fractionation signals could therefore complement observations of \ce{O2}--\ce{O2} for ocean loss planets, and these two sets of observations may be attainable with \textit{JWST}, at least for the inner planets of TRAPPIST-1. \citet{Lustig:2019} found that \ce{O2}-\ce{O2} features in \ce{O2}-dominated atmospheres may be detected in as few as ten transits (for TRAPPIST-1~b) in the outgassing case with the NIRSpec Prism {SUB512 ($n_\text{groups}=6$)}, equal to the ten transits required to {distinguish} HDO bands compared to VSMOW. 
The NIRSpec Prism is particularly useful because the extra wavelength coverage (0.6--5.3~\um{}) can provide simultaneous evidence of the \ce{O2} abundance (via \ce{O2}-\ce{O2} at 1.06 and 1.27~\um{}) and of isotopic fractionation of hydrogen in \ce{H2O} or oxygen in \ce{CO2}. Here we have shown that for a planet like TRAPPIST-1~b, the number of transits required to identify these individual \ce{O2}-\ce{O2} bands is sufficient to begin distinguishing isotopologues as well, without additional observing time than necessary to identify a large (abiotic) inventory of \ce{O2}.

{Although our motivation for assessing atmospheric and ocean loss through measuring the D/H ratio was inspired by remote-sensing measurements of D/H in the atmosphere of Venus, it will not likely be possible to conduct these measurements in the atmospheres of Venus-like exoplanets using transit observations. Sulfuric acid aerosol formation truncates the atmospheric levels that can be probed in transmission and the saturated \ce{CO2} bands largely blanket the NIR--MIR spectra of \ce{CO2}-dominated atmospheres. The effective greenhouses cause high temperatures, which disfavor the occupation of isotopologue ro-vibrational energy states \citep{Molliere:2019}.}

Isotopologue observations may help constrain the location of the inner edge of the habitable zone by providing evidence for whether water vapor detected in the atmosphere of a planet is consistent with ocean loss or with a primordial reservoir (i.e. surface water or outgassing). Since TRAPPIST-1~d sits between the recent Venus and moist greenhouse limits, it is a valuable target for this observation. The observation of water vapor with no evidence of fractionation in the atmosphere of an IHZ planet like TRAPPIST-1~d would support the recent Venus limit, though the lack of fractionation could also be due to outgassing fluxes or a high volatile inventory---i.e. it could be a water world in a permanent runaway greenhouse state, which would be indicated by a large abundance of statospheric water vapor. Conversely, isotopic evidence of ocean loss would strongly support the moist greenhouse limit as the true inner edge. Robust observational evidence for the location of the IHZ would require a survey of multiple planets and multiple systems with high-fidelity observations to constrain both the \ce{H2O} abundances and D/H ratios.

{It may be difficult to observe isotopologues in the atmospheres of planets within the habitable zone. We found that TRAPPIST-1~e exhibited only small signals, and even in the optimistic assessment we conducted that neglected any systematic noise floor, it would generally require greater than 100 transits to identify isotopologues at \SN{}=5. This is likely to apply to HZ planets in general, due to the lower atmospheric temperatures (and resultant lower atmospheric scale height), additional refraction due to distance from the star, and lower stratospheric water vapor.}

The fractionation we have modeled may not be attainable for all small planets that may have undergone ocean and atmospheric loss in and around the habitable zones of M~dwarf stars. If a planet targeted for observation is more volatile-rich than Earth, the volatile inventory available to the atmosphere may never be lost, and as discussed above, would not impart a significant fractionation signature on the atmosphere.  This may be the case for one or more of the TRAPPIST-1 planets, which may be more volatile rich than Earth due to slightly lower nominal densities \citep{Grimm:2018} and dynamical evidence of possible migration \citep{Luger:2017b}. However, within the calculated error, these densities are still within the range of terrestrial values {in a mass-radius relationship \citep{Grimm:2018}}.  

An observation of oxygen fractionation in \ce{CO2} could indicate extreme atmospheric loss but may be difficult to achieve.  The fractional mass difference between \ce{^18O} and \ce{^16O} is small, and Venus does not exhibit fractionation in oxygen \citep{Hoffman:1980,Bezard:1987,Iwagami:2015}. So significant fractionation in oxygen may require atmospheric escape mechanisms unknown in our solar system, or operating over longer time periods or at higher fluxes than for the solar system planets.  However, if it occurred, oxygen fractionation could be retained within \ce{CO2} if oxygen loss occurred in the presence of \ce{CO2}. Without surface liquid water, surface weathering processes \citep[i.e. carbonate-silicate weathering,][]{Walker:1981} would no longer draw outgassed \ce{CO2} from the atmosphere. Since the oxygen in \ce{CO2} would dilute the isotopic signal from oxygen loss, the remaining \ce{CO2} inventory must be small compared to the remaining oxygen to prevent significant dilution of the oxygen isotope abundances.   

In addition to potential dilution of oxygen isotopic fractionation, atmospheres with large inventories of \ce{CO2} would make it more difficult to observe any isotopic fractionation signal considered here.  Unlike the other typical terrestrial bulk atmospheric gases \ce{O2} and \ce{N2}, \ce{CO2} exhibits significant absorption throughout the NIR--MIR, which masks weaker absorption lines, including isotopologues. Large quantities of water vapor can similarly interfere with observing other trace gases. While Earth-like abundances of \ce{CO2} and \ce{H2O} may be the most favorable for distinguishing and characterizing important isotopologue bands, low \ce{CO2} abundance may not be likely for a planet that experienced total ocean loss because the lack of liquid surface water would reduce surface weathering that draws outgassed \ce{CO2} from the atmosphere.

{Other factors not considered in this work can affect the possibility of detecting isotopic fractionation, such as surface pressure and clouds. \citet{Morley:2017} showed that for atmospheres with surface pressure of 1 bar and lower, the number of transits required to detect features increased with decreasing pressure. In \citet{Lincowski:2018} the amplitudes of transit transmission features did not change significantly as a result of higher surface pressure at pressures higher than 10 bars. Clouds and hazes, which may form at high altitude, can significantly truncate transit transmission spectra, particularly hazes such as those in the atmosphere of Venus, which form at much higher altitude than water clouds \citep[e.g.][]{Lincowski:2018}. For Venus-like clouds in particular, \ce{H2SO4} has many absorption bands longward of 2.5~\um{}. These absorption bands are not the same as \ce{CO2} or \ce{H2O}, but together with these gases and due to the high altitude of haze aerosols, the presence of \ce{H2SO4} aerosols can effectively eliminate primary detectable isotopologue features in the 2.8--4.3~\um{} range, between \ce{CO2} bands. While TRAPPIST-1~d and e may form aerosols, \citet{Lincowski:2018} showed that, due to high atmospheric temperatures, {the most likely condensates in these oxidized atmospheres, \ce{H2O} and }\ce{H2SO4}, would not condense in the atmosphere of a Venus-like TRAPPIST-1~b.} {While other metal aerosols have been suggested for hotter exoplanets (such as sodium and potassium-based condensates), these are not plausible for the temperature regimes considered here, as it would not be possible to evaporate them from the planetary surface \citep[c.f.][]{Schaefer:2009}. Hydrogen-dominated atmospheres could support other types of aerosols, particularly if these planets reached higher temperatures \citep[e.g.][]{He:2018b}, but the TRAPPIST-1 planets are not likely to have H-dominated atmospheres \citep{DeWit:2016,DeWit:2018,Moran:2018}, with the possible exception of TRAPPIST-1~g \citep{Moran:2018}, which we do not consider here. Consequently, TRAPPIST-1b is not likely to support the majority of hypothesized aerosols, and so is more likely to be clear sky than other planets in the system.}

{Here TRAPPIST-1 b, d, and e were used as sample planets to assess the possibility of detecting isotopologue bands in exo-terrestrial atmospheres. The results could be extended to other planets in the system, in light of the results of \citet{Morley:2017}, \citet{Lincowski:2018}, and \citet{Lustig:2019}, or to other systems. While TRAPPIST-1~c is one of the inner planets with parameters similar to Venus, it is more difficult to observe features in the atmosphere of TRAPPIST-1~c than b and d, due to the higher density of c. Given the results for TRAPPIST-1~e, it is unlikely that isotopologue bands could be observed in the atmospheres of the outer planets, TRAPPIST-1~f, g and h. The results of this work could be used to assess the detectability of isotopologue bands in the atmospheres of other M~dwarf targets of interest, both those currently known and those yet to be discovered by {SPECULOOS \citep{SPECULOOS:2018}, TESS \citep{Barclay:2018}, CHEOPS \citep{Benz:2013,Benz:2018}, and PLATO \citep{Rauer:2014}.}}

We have shown that isotopic fractionation signals inferred from HDO or \OCO{} may be feasible to observe with \textit{JWST} and may provide important clues of the evolutionary history of planets around M~dwarfs. {Measurements with ground-based high-resolution instruments may also be able to search for isotopic signatures for the very nearest M~dwarf planets \citep{Molliere:2019}.} The values we have considered can guide observers considering different levels of fractionation. These levels could also be used to assess the possibility of detecting isotopic fractionation if future comprehensive modeling of atmospheric escape demonstrates at what level hydrogen or oxygen fractionation is possible, and serve as a testable hypothesis for fractionation due to ocean loss. After first determining that a planet has an atmosphere \citep{Lincowski:2018,Lustig:2019}, the next important assessment of the planetary environment may be to search for signs of severe ocean loss. Isotopic measurements can contribute evidence for assessing atmosphere and ocean loss of M~dwarf planets, which will soon be observed with \textit{JWST}.   Although severe atmospheric escape is likely to afflict all small planets in or near M~dwarf habitable zones, caveats against such fractionation occurring also exist. Observations of one or more inner planets in a multiple-planet  system could be used to inform the suitability of more time-consuming follow-up observations of a habitable zone sibling.

\section{Conclusions} \label{sec:conclusions}

We have shown that for Venus-like isotopic fractionation of D/H, or a similar fractionation in \ce{^18O/^16O}, isotopologue bands may be observable and distinguished from Earth-like isotopic abundances with \textit{JWST} in as few as ten transits (\ce{\delta{D}}=100) or four transits (\ce{\delta^18O}=100) for a {clear-sky} atmosphere not dominated by \ce{CO2}. The large fractionation values considered here are meant to demonstrate the potential for detecting these bands and discriminating them from Earth-like abundances. These fractionation values would require an ocean-free surface and ocean loss at least as severe as experienced by Venus. A detection of these bands in transit transmission spectra would be evidence of a lack of a surface ocean and the extreme atmospheric loss and oxygen build-up that has been proposed by a number of authors. This would provide valuable constraints for atmospheric escape models, the location of the inner edge of the habitable zone (whether recent Venus or moist greenhouse), and on the habitability of M~dwarf planets. Researchers currently preparing \textit{JWST} observing proposals, and those who will be conducting retrievals on future \textit{JWST} observations, may want to consider different isotopic abundances in line lists used as inputs to retrieval pipelines. Further work modeling atmospheric escape that includes elemental isotopes is also warranted, to understand to what degree isotopic fractionation is theoretically possible in the atmospheres of planets orbiting M~dwarfs. A thorough analysis of atmospheric escape should consider thermal and nonthermal escape, including photochemistry and vertical transport, life-long outgassing, and the possibility of deep surface reservoirs (i.e. water worlds).

\acknowledgements
We thank Rodrigo Luger and David Crisp for useful discussions that helped improve this paper. This work was performed as part of the NASA Astrobiology Institute's Virtual Planetary Laboratory, supported by the National Aeronautics and Space Administration through the NASA Astrobiology Institute under solicitation NNH12ZDA002C and Cooperative Agreement Number NNA13AA93A, and by the NASA Astrobiology Program under grant 80NSSC18K0829 as part of the Nexus for Exoplanet System Science (NExSS) research coordination network. A.P.L. acknowledges support from NASA Headquarters under the NASA Earth and Space Science Fellowship Program -- Grant 80NSSC17K0468. This work was facilitated though the use of advanced computational, storage, and networking infrastructure provided by the Hyak supercomputer system at the University of Washington. We also thank the anonymous reviewer, whose thoughtful comments helped us greatly improve the manuscript.

\software{SMART \citep{Meadows:1996,Crisp:1997}, LBLABC \citep{Meadows:1996}, PandExo \citep{Batalha:2017b}, GNU Parallel \citep{Tange:2011}}

\bibliographystyle{aasjournal} \bibliography{linc}

\end{document}

%% file: table1.tex
\begin{table*}
{\iftwocol
    \footnotesize\selectfont
\else
    \scriptsize\selectfont
\fi
\centering
\caption{{Simulated Environments  \label{table:environments}}}
\begin{tabular}{llll}
\hline \hline
Planetary State            & Key Gases                 & D/H     & \ce{^18O}/\ce{^16O} \\ \hline
\ce{O2} outgassing              & 95\% \ce{O2}, trace \ce{H2O}, \ce{CO2}, CO, \ce{O3}     & 10, 100 & 1, 10   \\
\ce{O2} outgassing, reduced \ce{H2O} & 95\% \ce{O2}, trace \ce{H2O}, \ce{CO2}, CO, \ce{O3} & 10, 100 & 1, 10   \\
\ce{O2} desiccated              & 95\% \ce{O2}, 4.5\% \ce{N2}, 0.5\% \ce{CO2}, trace CO, \ce{O3}       & -       & 10, 100 \\
Venus-like clear-sky       & 96.5\% \ce{CO2}, 3.5\% \ce{N2}, trace \ce{H2O}, CO  & 10, 100 & 1, 10  \\ \hline
\end{tabular}
\flushleft
  {\textbf{Note:} Climatically and photochemically self-consistent environments from \citet{Lincowski:2018} simulated for spectral analysis, dominated either by \ce{O2} or \ce{CO2}, assuming a range of trace species outgassing. TRAPPIST-1~b, d, and e were simulated for all environments with the listed fractionation levels. We model a reduced \ce{H2O} case for the \ce{O2} outgassing environments, where the water vapor is scaled down by a factor of 100 (TRAPPIST-1~b) or 10 (TRAPPIST-1 d), which reduces the water vapor to the level observed in the stratospheres of Venus and Earth ($\sim1-3$~ppm). As a result, the reduced \ce{H2O} atmospheres do not have self-consistent climates/photochemistry. There was no reduced \ce{H2O} case for TRAPPIST-1~e because the water vapor was already $\sim$1--5~ppm. }\\}
\end{table*}

%% file: table2.tex
\begin{table*}
{\iftwocol
    \footnotesize\selectfont
\else
    \scriptsize\selectfont
\fi
\centering
\caption{Calculated Isotopologue Abundances  \label{table:abundances}}
\begin{tabular}{rlllllllll}
\hline \hline
\ce{\delta D}, \ce{\delta^18O} & VSMOW     & 1, 2     & 1, 10    & 1, 100   & 10, 1    & 10, 2    & 100, 1   & 100, 2   & 100, 10  \\
                 \hline
\ce{CO2}          & 0.984               & 0.979               & 0.938               & 0.577               & 0.984               & 0.979               & 0.984               & 0.979               & 0.938               \\
\ce{^13CO2}       & 0.011               & 0.011               & 0.011               & 6.48$\times10^{-3}$ & 0.011               & 0.011               & 0.011               & 0.011               & 0.011               \\
\ce{^16O^12C^18O} & 3.95$\times10^{-3}$ & 7.87$\times10^{-3}$ & 0.039               & 0.313               & 3.95$\times10^{-3}$ & 7.87$\times10^{-3}$ & 3.95$\times10^{-3}$ & 7.87$\times10^{-3}$ & 0.039               \\
\ce{^16O^12C^17O} & 7.34$\times10^{-4}$ & 2.49$\times10^{-3}$ & 0.012               & 0.099               & 7.34$\times10^{-4}$ & 2.49$\times10^{-3}$ & 7.34$\times10^{-4}$ & 2.49$\times10^{-3}$ & 0.012               \\
\ce{^16O^13C^18O} & 4.43$\times10^{-5}$ & 8.84$\times10^{-5}$ & 4.33$\times10^{-4}$ & 3.51$\times10^{-3}$ & 4.43$\times10^{-5}$ & 8.84$\times10^{-5}$ & 4.43$\times10^{-5}$ & 8.84$\times10^{-5}$ & 4.33$\times10^{-4}$ \\
\ce{^16O^13C^17O} & 8.25$\times10^{-6}$ & 2.79$\times10^{-5}$ & 1.37$\times10^{-4}$ & 1.11$\times10^{-3}$ & 8.25$\times10^{-6}$ & 2.79$\times10^{-5}$ & 8.25$\times10^{-6}$ & 2.79$\times10^{-5}$ & 1.37$\times10^{-4}$ \\
\ce{C^18O2}       & 3.96$\times10^{-6}$ & 9.23$\times10^{-6}$ & 5.07$\times10^{-5}$ & 4.19$\times10^{-4}$ & 3.96$\times10^{-6}$ & 9.23$\times10^{-6}$ & 3.96$\times10^{-6}$ & 9.23$\times10^{-6}$ & 5.07$\times10^{-5}$ \\
\ce{^17O^12C^18O} & 1.47$\times10^{-6}$ & 5.83$\times10^{-6}$ & 3.20$\times10^{-5}$ & 2.65$\times10^{-4}$ & 1.47$\times10^{-6}$ & 5.83$\times10^{-6}$ & 1.47$\times10^{-6}$ & 5.83$\times10^{-6}$ & 3.20$\times10^{-5}$ \\
\ce{C^17O2}       & 1.37$\times10^{-7}$ & 9.22$\times10^{-7}$ & 5.06$\times10^{-6}$ & 4.19$\times10^{-5}$ & 1.37$\times10^{-7}$ & 9.22$\times10^{-7}$ & 1.37$\times10^{-7}$ & 9.22$\times10^{-7}$ & 5.06$\times10^{-6}$ \\
\ce{^13C^18O2}    & 4.45$\times10^{-8}$ & 1.04$\times10^{-7}$ & 5.69$\times10^{-7}$ & 4.71$\times10^{-6}$ & 4.45$\times10^{-8}$ & 1.04$\times10^{-7}$ & 4.45$\times10^{-8}$ & 1.04$\times10^{-7}$ & 5.69$\times10^{-7}$ \\
\ce{^18O^13C^17O} & 1.65$\times10^{-8}$ & 6.55$\times10^{-8}$ & 3.60$\times10^{-7}$ & 2.98$\times10^{-6}$ & 1.65$\times10^{-8}$ & 6.55$\times10^{-8}$ & 1.65$\times10^{-8}$ & 6.55$\times10^{-8}$ & 3.60$\times10^{-7}$ \\
\ce{^13C^17O2}    & 1.54$\times10^{-9}$ & 1.04$\times10^{-8}$ & 5.69$\times10^{-8}$ & 4.70$\times10^{-7}$ & 1.54$\times10^{-9}$ & 1.04$\times10^{-8}$ & 1.54$\times10^{-9}$ & 1.04$\times10^{-8}$ & 5.69$\times10^{-8}$ \\
\hline
\ce{H2O}          & 0.997               &                     &                     &                     & 0.995               & 0.992               & 0.967               & 0.964               & 0.944               \\
\ce{H2 ^18O}      & 2.00$\times10^{-3}$ &                     &                     &                     & 1.99$\times10^{-3}$ & 3.98$\times10^{-3}$ & 1.94$\times10^{-3}$ & 3.87$\times10^{-3}$ & 0.019               \\
\ce{H2 ^17O}      & 3.72$\times10^{-4}$ &                     &                     &                     & 3.71$\times10^{-4}$ & 1.26$\times10^{-3}$ & 3.61$\times10^{-4}$ & 1.22$\times10^{-3}$ & 5.99$\times10^{-3}$ \\
\ce{HDO}          & 3.11$\times10^{-4}$ &                     &                     &                     & 3.10$\times10^{-3}$ & 3.09$\times10^{-3}$ & 0.031               & 0.031               & 0.030               \\
\ce{HD^18O}       & 6.23$\times10^{-7}$ &                     &                     &                     & 6.22$\times10^{-6}$ & 1.24$\times10^{-5}$ & 6.14$\times10^{-5}$ & 1.22$\times10^{-4}$ & 5.99$\times10^{-4}$ \\
\ce{HD^17O}       & 1.16$\times10^{-7}$ &                     &                     &                     & 1.16$\times10^{-6}$ & 3.92$\times10^{-6}$ & 1.14$\times10^{-5}$ & 3.87$\times10^{-5}$ & 1.89$\times10^{-4}$ \\
\ce{D2O}          & 2.42$\times10^{-8}$ &                     &                     &                     & 3.45$\times10^{-7}$ & 3.44$\times10^{-7}$ & 3.53$\times10^{-6}$ & 3.52$\times10^{-6}$ & 3.45$\times10^{-6}$ \\
\hline

\end{tabular}
\flushleft
  \textbf{Note:} The column headers list each pair of hydrogen and oxygen isotope fractionation values compared to Vienna Standard Mean Ocean Water (VSMOW, from HITRAN2016; \citealp{HITRAN:2016}).{ Note VSMOW D/H=$1.5574\times10^{-4}$ \citep{Abundances:2016}.} Not listed here are adjusted abundances for CO, \ce{O2}, and \ce{O3}, which we include in our models. The primary potentially observable species are HDO and \OCO{}.\\}
\end{table*}